\documentclass[prd, notitlepage, twocolumn, nofootinbib]{revtex4-1}
\usepackage{graphicx}
\usepackage{amsmath,mathtools}
\usepackage[colorlinks]{hyperref}
\usepackage[caption=false]{subfig}

\newcommand{\tb}{t_\beta}

\newcommand{\cba}{c_{\beta-\alpha}}
\newcommand{\sba}{s_{\beta-\alpha}}

\newcommand{\mhp}{m_{H^\pm}}

\newcommand{\fb}{\text{ fb}}
\newcommand{\gev}{\text{ GeV}}
\newcommand{\tev}{\text{ TeV}}
\newcommand{\hzg}{Z H_i \gamma}
\newcommand{\theprocess}{e^+e^- \rightarrow \hzg}

\newcommand{\processBir}{e^+e^- \rightarrow Z h^0 \gamma}
\newcommand{\processIki}{e^+e^- \rightarrow Z H^0 \gamma}
\newcommand{\processUc}{e^+e^- \rightarrow Z A^0 \gamma}

\begin{document}



\title{Production of neutral Higgs bosons associated with Z-boson and photon in 2HDM at future lepton colliders}

\author{Nasuf Sonmez}
\email{nasuf.sonmez@ege.edu.tr}
\affiliation{Ege University, Faculty of Science, Physics Department, Izmir 35040, Turkey}

\date{\today}

\begin{abstract}
Inspired by precision tests of the Standard Model in future lepton colliders, the numerical analysis of the following scattering processes, $e^+e^- \rightarrow  Z h^0\gamma$ and $e^+e^- \rightarrow Z H^0 \gamma$, are carried at the tree level including all possible diagrams in the two-Higgs-doublet model (2HDM). This model has many free parameters, but the parameters which take part in the scattering amplitudes of these two processes are primarily the mixing angle parameter, $s_{\beta-\alpha}$, and the masses of the neutral Higgs bosons, ($h^0, H^0$). Therefore, measuring the production rates of  $Z h^0\gamma$ and $Z H^0 \gamma$  final states open another test for the scalar sectors of the 2HDM. The numerical analysis is performed under the current experimental constraints. The production rates and the asymmetry in the forward-backward direction are presented as a function of the center-of-mass energy covering the future lepton colliders.  The unpolarized cross-section  gets up to 6.19 (4.86) fb at $\sqrt{s} = 350$ (500) GeV and 0.164 (0.157) fb at $\sqrt{s} = 350$ (500) GeV for  $e^+e^- \rightarrow  Z h^0\gamma$ and $e^+e^- \rightarrow Z H^0 \gamma$, respectively. The polarization of the incoming $e^+e^-$ beams are studied for various configurations, and it enhances the cross-section by a factor of 1.78 in both processes for $P_{e^+,e^-}=(+0.6,-0.8)$.
\end{abstract}

\maketitle

\section{Introduction}
\label{sec1}

The last missing part namely Higgs boson, which is related with the electroweak symmetry breaking mechanism in the Standard Model (SM), was discovered at the LHC \cite{Aad:2012tfa, Chatrchyan:2012xdj, Khachatryan:2014jba}. The production and the decay channels of the Higgs particle have been studied extensively since then \cite{Aad:2015gba, 2015PhRvL.114s1803A}. The discovery commenced questioning which model it belongs to. Many studies for its properties have been reported, and experiments are still continuing. All the results presented by ATLAS and CMS collaborations showed that the findings resemble a Higgs boson in the SM. 
It is possible that the SM is a low energy approximation of a larger theory which explains up to the Fermi scale. There are many proposals which extend the SM, and they have been studied extensively in the last decades. A simple extension of the SM is to enlarge only the scalar sector of the model, which is called the two-Higgs-doublet model; mainly, a new Higgs doublet is added to the theory.
So that these two Higgs doublets can couple to matter and gauge bosons and give mass to leptons, quarks, and electroweak bosons. In the 2HDM, there are two charged Higgs bosons $(H^\pm)$ and three neutral Higgs bosons $(h^0,A^0,H^0)$ \cite{Branco:2011iw, Gunion:1989we}. The extra doublet generates new couplings and interactions. As a result, rich phenomenology arises, especially for Higgs physics.

To be able to unravel the properties of this particle, the collision of fundamental particles other than protons is needed. 
The best candidate is the lepton, and a lepton collider is an excellent machine for studying the properties of the Higgs boson.
Various proposals for future lepton colliders have been submitted around the world running at the center of mass energies between 240 - 1000 GeV; 
they are the Circular Electron-Positron Collider (CEPC) in China \cite{CEPCStudyGroup:2018rmc, CEPCStudyGroup:2018ghi}, 
the Future Circular Collider (FCC-ee) \cite{Benedikt:2651299} at CERN \cite{Gomez-Ceballos:2013zzn}, 
and the International Linear Collider (ILC) in Japan \cite{Yamamoto:2017lnu}. 
In all these proposals, the planned facilities will produce millions of Higgses by electron-positron collisions. Thus, the properties of these Higgses can be revealed with high statistical precision.

Two of the well-motivated channels at the $e^+e^-$-colliders are the production of Higgses with gauge bosons and fermions, namely the production of $ZH$ and $H\nu_e\bar{\nu}_e$. 
These channels will help to determine the couplings, $c_{HZZ}$  and  $c_{HWW}$, and the Higgs total decay width \cite{Thomson:2015jda, McCullough:2013rea, Rao:2006hn, Sun:2016bel, Yan:2016xyx, Chen:2016zpw, 2018arXiv180604992Z}. 
Studying double Higgs-strahlung along with WW double-Higgs fusion makes it possible to determine the triple Higgs self-coupling ($c_{HHH}$) with an astonishing precision \cite{Djouadi:1999gv, GutierrezRodriguez:2008nk, Battaglia:2001nn, dEnterria:2016sca, Castanier:2001sf}. Measuring Higgs self-couplings is essential for reconstructing the Higgs potential in the SM. The complete reconstruction of the Higgs potential in the SM requires the determination of the quartic Higgs self-coupling $c_{HHHH}$ as well. This coupling can be reached directly in the triple Higgs production \cite{Djouadi:1999gv, Sonmez:2018smv}. 
There are some couplings which are not defined at tree-level but arise at loop-level in the SM, and they are called the anomalous trilinear Higgs couplings, $HZ\gamma$, $H\gamma\gamma$ and $HZZ\gamma$. These couplings are sensitive to new physics contributions through new massive particles propagating in the loops \cite{Cao2013, Han2013}. In all these processes, the couplings of Higgses with gauge bosons are significant to determine the other Higgs couplings. Since there are additional Higgs states in the 2HDM, the determination of the Higgs couplings becomes entangled. 

Many authors have studied the scattering process $e^+e^-\rightarrow ZH$, and measuring the cross-section is also one of the objectives which are shared among all three future colliders. Besides this channel,  the $ZH$ could be produced in association with a high energetic photon. The process $e^+e^-\rightarrow ZH\gamma$ suits great for studying the couplings of the Higgs to neutral gauge bosons in the SM, and it is also part of the inclusive process $e^+e^-\rightarrow Z H+X$. There are three neutral Higgs bosons in the 2HDM, so there are three similar processes: $\processBir$ , $\processIki$, and $\processUc$. The relevant Higgs couplings for these processes are  $c_{\{h^0, H^0\}ZZ}$ and $c_{\{h^0, H^0\} A^0Z}$, therefore, the situation is complicated compared to the SM. 
Additionally, the couplings $c_{h^0ZZ}$ and $c_{H^0ZZ}$ in the 2HDM influence the anomalous trilinear couplings. 
Compared to the $Z H$ channel, the production cross-section of  $\theprocess$ is considerably less, but they still have a moderate cross-section, and it is possible to study them in the colliders. It could be considered as another test for the Higgs couplings with an additional vector boson (photon) at the final state. 
The production of Higgs boson with additional vector bosons turns into a complicated problem in extended Higgs models. Therefore, the following questions need to be answered:  "Which couplings are significant for these scattering processes?", "What are the relevant free parameters that take place in each scattering process?", moreover, "What could be extracted from these scattering processes in the context of 2HDM?". 

The scattering process $e^+e^-\rightarrow ZH\gamma$ in the SM was calculated before in ref. \cite{Liu:2013cla} at the loop level. The same process was studied in ref. \cite{Alam:2017hkf} by taking into account the CP-conserving couplings between Higgs to gauge bosons within the effective Lagrangian framework and dimension-six operators, and the effects of these anomalous Higgs-gauge boson couplings were compared. 
This paper presents the calculation of $\theprocess$ in the 2HDM, including all possible diagrams. The cross-section is calculated as a function of the center-of-mass (CM) energy, and comparison is carried out regarding the proposed colliders.
Dependence of the cross-sections on the free parameters is presented. Several polarization configurations of the $e^+e^-$ beams are assumed. The forward-backward asymmetry between the Higgs boson and the photon is calculated, and a discussion is carried for the colliders assumed.

The layout of this paper is organized as follows. In section \ref{sec2}, the theory of the 2HDM, all constraints, and assumptions in the computation are discussed. In section \ref{sec3}, expressions regarding the kinematics of the scattering and the total cross-section with polarized electron-positron beams are emphasized. In section \ref{sec4}, numerical results of the total cross-section with polarization and the asymmetry distributions are presented. The conclusion is drawn in section \ref{sec5}.

\section{Theoretical framework of the 2HDM}
\label{sec2}

The 2HDM has been studied before, and a detailed introduction of the 2HDM framework was given by various authors \cite{Branco:2011iw, Gunion:1989we, Haber:2006ue, Davidson:2005cw, Carena:2002es}. Therefore, we only give a summary of the 2HDM, particularly the scalar potential and the free parameters in the model. The 2HDM is constructed by adding a second $SU(2)_L$ Higgs doublet with the same hypercharge ($Y=1$) to the scalar sector. Imposing a discrete symmetry, $\mathcal{Z}_2$, on the scalar potential helps to avoid flavor-changing-neutral-currents (FCNC) \cite{Glashow:1976nt} at the tree-level. In this study, CP-conservation is considered. Accordingly, the scalar potential is defined as follows:
\begin{eqnarray}
        V(\Phi_1,\Phi_2)&=&m_{11}^2 | \Phi_1|^2+m_{22}^2|\Phi_2|^2 - \left[ m_{12}^2   \Phi_1^{\dagger} \Phi_2 +h.c. \right] \nonumber \\ 
        &+& \frac{\lambda_1}{2}| ( \Phi_1^{\dagger} \Phi_1 )^2  + \frac{\lambda_2}{2} ( \Phi_2^{\dagger} \Phi_2 )^2 +\lambda_3 | \Phi_1 |^2 | \Phi_2 |^2 \label{eq:eq2}  \nonumber \\
        &+&\lambda_4 | \Phi_1^{\dagger} \Phi_2 |^2 + \left [  \frac{\lambda_5}{2} (\Phi_1^{\dagger} \Phi_2)^2  + h.c.  \right] \;,
        \end{eqnarray}
where the coupling constants, $\lambda_i, (i=1,..,4)$, are real, the parameters $m_{12}^2$ and $\lambda_{5}$ can be complex, but they are taken real for simplicity, and $\Phi_i,\;(i=1,2)$ are the Higgs doublets.

Following the prescription defined in ref. \cite{Kanemura:2015ska} the masses of all the extra Higgs bosons can be calculated as usual. First, the stationary conditions of the potential are applied, and several constraints are obtained. Second, these bound conditions are substituted into the scalar potential to eliminate $m_{11}$ and $m_{22}$. Then the scalar potential decomposes into a quadratic term plus cubic and quartic ones \cite{Haber:1978jt, Branco:2005jr, Haber:2006ue}. Finally, diagonalizing the quadratic mass terms, the physical Higgs states and their masses are obtained. The other terms, cubic and quartic ones, define the couplings and the interactions in the model. 
As a result, the free parameters of the model are the masses of the neutral Higgs bosons $(m_{h/H^0/A^0})$ and charged Higgs bosons ($m_{H^\pm}$), the ratio of the vacuum expectation values ($\tb=\tan\beta=v_1/v_2$), the mixing angle between the CP-even neutral Higgs states ($\alpha$) and the soft breaking scale of the discrete symmetry $m^2=m_{12}^2/(\sin\beta\cos\beta)$ \cite{Gunion:2002zf}. The masses of charged and CP-odd Higgs states are defined as follows:
        \begin{equation}
        m^2_{A^0}   =\frac{m_{12}^2}{\sin\beta\cos\beta} -\lambda_5 v^2, \;\;     m^2_{H^\pm} =m^2_{A^0} + \frac{1}{2}(\lambda_5-\lambda_4)v^2\;.
        \end{equation}
The masses of the CP-even states are defined below.
        \begin{equation}
        \begin{pmatrix}
        h^0 \\
        H^0        
        \end{pmatrix} 
        = \mathcal{R}
        \begin{pmatrix}
        m_{12}^2\tb +\lambda_1 v_1 ^2        &    -m_{12}^2+\lambda_{345} v_1 v_2    \\
        -m_{12}^2+\lambda_{345} v_1 v_2 & m_{12}^2/\tb +\lambda_2 v_2^2
        \end{pmatrix} \mathcal{R}^T
        \end{equation}
where $\mathcal{R}$ is a unitary rotation matrix which diagonalizes the CP-neutral Higgs mass matrix \cite{Davidson:2005cw} as a function of $(\beta-\alpha)$, and $\lambda_{345}=\lambda_3+\lambda_4+\lambda_5$.  After the discovery of the Higgs boson at the LHC, $h^0$ state was extensively defined as the SM-like Higgs boson with the same mass and couplings. This choice was called the exact alignment limit where $(\beta-\alpha)=\pm\pi/2$. The angle $\sba=\sin(\beta-\alpha)$ defines the mixing between the CP-even Higgs states. In this study, the exact alignment limit is chosen, $\sba=1$, consequently, $h^0$ becomes indistinguishable from the SM Higgs boson $H$.

To suppress the FCNC at the tree-level in the Higgs-fermion interactions, the same discrete symmetry, which was imposed before on the scalar potential, can be extended to Yukawa sector. These interactions are written in four different and independent ways. The couplings between fermions and Higgses are named Type-I to -IV\cite{Branco:2011iw}. Setting a different Yukawa coupling scheme does not have a noticeable effect on the results. 
Therefore, the calculation is performed only considering the Type-I Yukawa coupling scheme, though the results are the same in the numerical precision for Type-II.

\section{Machinery for the numerical analysis and the parameter space}
\label{sec3}

\subsection{Scattering processes and the cross-section}
The machinery of the calculation and conventions of the scattering processes are presented here. Throughout this paper, the scattering processes are denoted as
    \begin{eqnarray}
    e^+ (k_1,\mu) +e^- (k_2,\nu)\;\rightarrow \; Z (k_3) + H_i (k_4)+ \gamma (k_5),
    \label{eq:eq3}
    \end{eqnarray}
where $k_a$ $(a=1,...,5)$ are the four-momenta of  the incoming positron and electron beam, the outgoing Z-boson, neutral Higgs-bosons, and photon. Additionally, spin polarizations of the positron and the electron are indicated by $\mu$ and $\nu$, respectively. Feynman diagrams which contribute to the process $\theprocess$ at the tree-level are shown in figure \ref{fig1}. The 2HDM Lagrangian and a detailed phenomenological discussion were reported in refs. \cite{Haber:2006ue, Davidson:2005cw, Gunion:1989we, Carena:2002es}. The vertices were defined in \textsc{FeynArts} \cite{Kublbeck:1992mt, Hahn:2000kx}; the amplitudes are constructed using \textsc{FeynArts}. 
Since the computation is carried out at the tree-level, the \emph{Unitary gauge} suits better for the calculation. The simplification of the fermion chains, squaring amplitudes, and the numerical analysis are accomplished using the driver program given in \textsc{FormCalc} \cite{Hahn:2006qw}.

\begin{widetext}
\onecolumngrid
\begin{figure}[htbp]
\centering
\includegraphics[width=.75\textwidth]{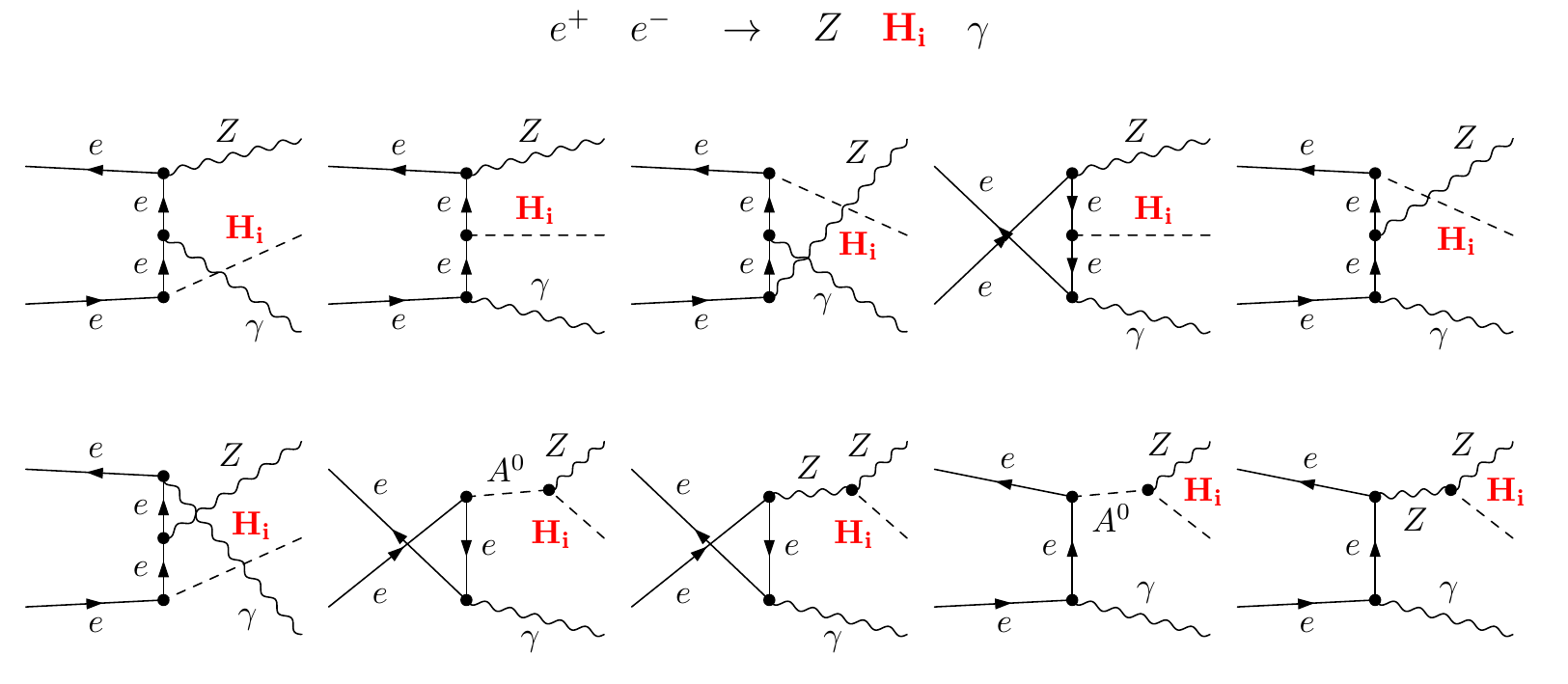}
\caption{\label{fig1} The Feynman diagrams which contribute each of the scattering processes at the tree level are plotted. $H_i$ represents the neutral Higgs bosons $h^0$, $H^0$, and $A^0$. In the production of $ZA^0\gamma$, the intermediated propagators $A^0$ and $Z$ are exchanged with $h^0$ and $H^0$ boson, respectively.}
\end{figure}
\end{widetext}

The relevant couplings for all the scattering processes are given in Table \ref{tab1}. The parameters $Y_{(1,2,3)}=(c_\alpha/s_\beta,s_\alpha/s_\beta,-\cot\beta)$ for Type-I and -IV, and they are $(-s_\alpha/c_\beta, c_\alpha/c_\beta,\tan\beta)$ for Type-II and -III. 
It can be seen that the most important parameter for all the scattering processes is $\sba$. It is involved in  $c_{\{h^0,H^0\}A^0Z}$ and $c_{\{h^0,H^0\}ZZ} $ couplings. Therefore, these couplings are the effective ones in each scattering process.
The couplings between the incoming electron-positron and the neutral Higgs bosons, $c_{e^+e^- \{h^0, H^0, A^0\}}$, are also a function of the free parameters of the 2HDM through $Y_i$. However, the factor $m_e/m_w$ suppresses them heavily; thus, they do not contribute significantly. As a result, it can be said that choosing one Yukawa coupling scheme (Type-I to -IV) over the others makes no important difference in all the scattering processes, and thus the total scattering amplitudes are affected only by the mixing angle $\sba$.

\begin{table}[htbp]
\centering 
\caption{\label{tab1} The couplings involved in all scattering processes. The weak angle is abbreviated as $s_w=\sin\theta_w$, $c_w=\cos\theta_w$, and $\cba=\cos(\beta-\alpha)$.}
\footnotesize
\begin{tabular*}{80mm}{l@{\extracolsep{\fill}}l | lc}
\hline\hline
$c_{\{\gamma,Z\}e^+e^-}$        & & $i e\left \{
\begin{pmatrix}        1 \\        1\\         \end{pmatrix},
\begin{pmatrix}        c_{2w}/s_{2w} \\   -s_w/c_w\\  \end{pmatrix}\right \}$                                \\
$c_{\{h^0,H^0,A^0\}e^+e^- }$  & & $-\frac{i e m_e }{2m_w s_w} \left\{ 
\begin{pmatrix}        Y_1 \\        Y_1\\         \end{pmatrix},
\begin{pmatrix}        Y_2 \\        Y_2\\         \end{pmatrix},
\frac{1}{i}\begin{pmatrix}        Y_3 \\        -Y_3\\         \end{pmatrix} \right\}$        \\
$c_{\{h^0,H^0\}A^0Z}$     & & $\frac{e}{2c_w s_w}\{\cba,\sba,\}$                                           \\
$c_{\{h^0,H^0\}ZZ  } $      & & $\frac{i e m_w}{c_w^2 s_w}\{\sba,\cba\}$                          \\
\hline
\hline
\end{tabular*}
\end{table}

\begin{figure}[htbp]
\centering
\includegraphics[width=0.48\textwidth]{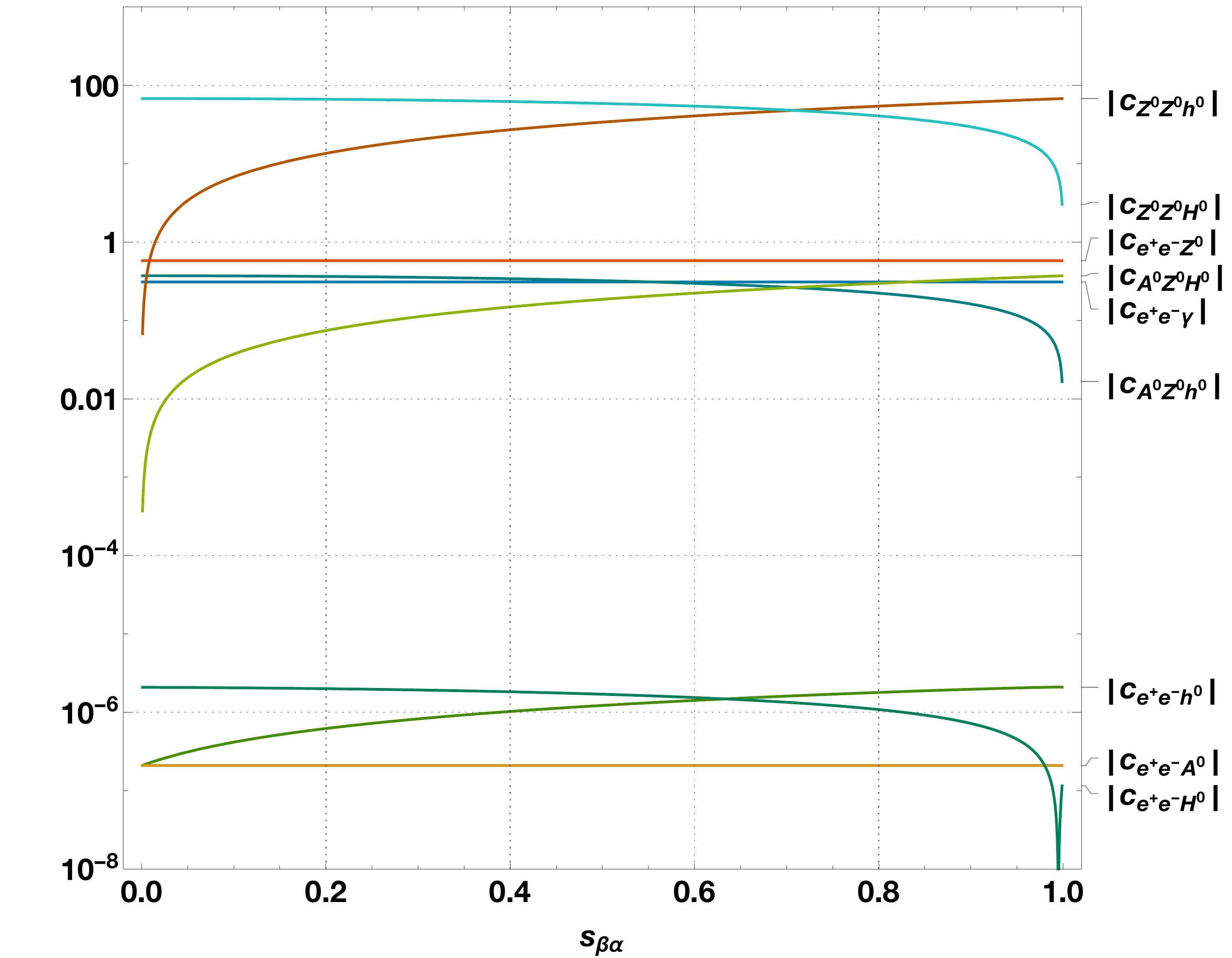}
\caption{\label{fig2} The couplings which take part in the scattering of all the channels are plotted as a function of $\sba$. The $\tb=10$ and Type-I Yukawa coupling scheme are assumed.}
\end{figure}

The couplings given in Table \ref{tab1} are plotted in Figure \ref{fig2} as a function of $\sba$. In Figure \ref{fig2}, Type-I Yukawa coupling scheme and $\tb=10$ are assumed. Though the numerical values of the couplings are the same for Type-I, only the couplings $c_{e^+e^- \{h^0,H^0,A^0\}}$ get lower. 
The highest contribution comes from $c_{ZZh^0}$ or $c_{ZZH^0}$ couplings, at the order of 65.
The coupling $c_{ZZH^0}$ has the maximum value at $\sba=0$, and it lowers when $\sba\rightarrow1$, while the coupling $c_{ZZh^0}$ rises with $\sba$. It should be noted that there is no $c_{ZZA^0}$ coupling in the 2HDM, and thus the cross-section of $\processUc$ is suppressed heavily. 
The couplings between fermion pairs and the neutral Higgses are at the order of $10^{-5}$, so they do not have much impact on the total amplitude. The couplings $c_{e^+e^-\gamma}$ and $c_{e^+e^-Z}$ are universal, and they are not a function of the free parameters of the 2HDM. Last, the couplings $c_{A^0 Z \{h^0, H^0\}}$ are also a function of $\sba$, but their values are around $0.37$; one of them drops while the other one rises. In conclusion,  the couplings $c_{ZZh^0}$ and  $c_{ZZH^0}$ are the most effective ones, and the cross-section of all these three processes are influenced heavily by $\sba$. 

The differential cross-section for three final state is defined as follows \cite{Kublbeck:1990xc}:

\begin{equation}
\frac{d\sigma (s;\mu,\nu)}{dk^0_5 dk^0_3 d\cos\theta \, d\eta}=\frac{2}{ (4\pi)^4}\frac{1}{\Phi} \left(\frac{1}{4} \sum_{pol}{|\mathcal{M}(s;\mu,\nu)_{tot}|^2}\right),
\label{eq:partcross}
\end{equation}
where the amplitude $\mathcal{M}_\text{tot}$  is squared and summed over the polarization vectors, $\Phi=2\sqrt{\lambda(s,m_e^2,m_e^2)}$ is the flux factor of the incoming $e^+e^-$ beams. The Monte-Carlo integration methods are required in the computation over the phase space of the final states, therefore, \textsc{CUBA} \cite{Hahn:2005pf, Hahn:2016ktb} routines are used. The cross-section for an arbitrary degree of longitudinal beam polarization is defined as follows:
\begin{equation}
    \label{eq:eqpolcross}
    \sigma(P_{e^+},P_{e^-})=\frac{1}{4} \sum_{\alpha,\beta=\pm1} (1+\alpha P_{e^+})(1+\beta P_{e^-}){\sigma}_{\alpha\beta} \;,
\end{equation}
The $\sigma_{LR}$ in equation \ref{eq:eqpolcross} represents the cross-section with a completely left-handed polarized ($P_{e^+} = -1=L$) positron beam and a right-handed polarized electron beam ($P_{e^-} = +1=R$). The cross-sections $\sigma_{RL}$, $\sigma_{LL}$ and $\sigma_{RR}$ are defined analogously. Note that due to the nature of the scattering process $\sigma_{LL}$ and $\sigma_{RR}$ don't make any important contribution, and they are neglected safely in equation \ref{eq:eqpolcross}.

Finally, due to the massless nature of the photon and the phase space of the outgoing particles, soft and collinear IR singularities appear naturally in the calculation. Therefore, motivated from photon energy resolution of the experimental apparatus (detector), we applied cut on the transverse momentum ($p_{T}^\gamma$) of the photon. As a result, the cross-section is calculated in a realistic environment. The cross-sections of each process are explored for varying the transverse momentum of the photon which is required to be $p_{T}^\gamma>p_{T,\text{cut}}^\gamma=(5,10,15,20,25)\gev$.

\subsection{Constraints and benchmark scenario}
The first set of constraints comes from a theoretical point of view: stability, unitarity, and perturbativity. Stability ensures that the scalar potential is positive even at large values of the field \cite{el2007consistency,  sher1989electroweak, deshpande1978pattern, Nie:1998yn, ElKaffas:2006gdt}. Unitarity ensures that the scattering amplitudes are flat at asymptotically high energies \cite{ginzburg2005tree}. Perturbativity is the condition where all the quartic couplings in the 2HDM need to be smaller than a particular value ($16\pi$). In this study, \textsc{2HDMC}(v1.7.0) \cite{Eriksson:2009ws} is used to test  whether the benchmark points obey these theoretical constraints.

The 2HDM was studied by many in the previous and still ongoing experiments, and numerous results have been piled up. All these results restrict the parameter space of the 2HDM from various angles. It was concluded in the previous section that the amplitudes are dependent mostly on the $\sba$. The phase spaces in all the scattering processes are a function of the masses of the neutral Higgs bosons. Therefore, all the scattering processes are unaffected by all the experimental constraints, except the $\sba$ and the masses of the neutral Higgs bosons.
Charged currents are important for the flavor observables because they can make novel contributions. However, we do not regard the charged Higgs mass constraints, because it does not make any direct contribution to the process (see figure \ref{fig1} for couplings and intermediated particles). However, the value of $\mhp$ most definitely affects the couplings in the model. Choosing up a wrong mass for the charged Higgs boson may not hold the theoretical and experimental constraints introduced in this section.

Inspired by ref. \cite{Enomoto:2015wbn} and the results presented by LHC \cite{Moretti:2016qcc}, we set masses of the extra Higgs bosons as $m_{A^0} = m_{H^\pm}$. This also satisfies the constraints on the oblique parameters \cite{Peskin:1990zt, Peskin:1991sw, Haber:2010bw, gunion2003cp, Grimus:2008nb, Polonsky:2000rs}. 
It should be underlined that if the parameter $\sba$ is fixed to unity, then the couplings of light Higgs ($h^0$) resemble the discovered SM Higgs boson. 
The parameter $\tb$ affects the Yukawa couplings, but they make no important contribution to the scattering processes; still, a value needs to be assigned. 
According to ref. \cite{Enomoto:2015wbn}, the low $\tb$ values were strongly constrained by $\bar{\mathcal{B}}(B_s^0 \rightarrow \mu^+\mu^-)$ and neutral meson mixings $\Delta M_s$ in Type-I, so $\tb>1$ can be considered. The masses of the scalars are not constrained on large $\tb$ range compared to the Type-II. Thus, $\tb=10$ is picked.
The calculation is performed following those experimental constraints.

\section{Results and discussion}
\label{sec4}

In this section, the numerical results for the production of $\hzg$ in an $e^+e^-$-collider are presented and discussed. The following SM parameters are taken from ref. \cite{Tanabashi:2018oca}: $m_e=0.5109989\text{ MeV}$, $m_Z=91.1876\gev$,   $m_{W^\pm}=80.385\gev$ 
and $\alpha=1/127. 944$. We used the current known value for the Higgs mass $m_H=125.09\gev$ \cite{2015PhRvL.114s1803A}.

\subsection{Cross-section distributions}

The numerical analysis is carried out for unpolarized and polarized incoming beams. If it is not stated otherwise, we set $m_{H_i}=150\gev$ and $\tb=10$ in the computation. In figure \ref{fig3}, the cross-section distributions are given for various polarizations as a function of the CM energy. These polarization configurations were discussed before in various refs. such as \cite{Fujii:2018mli,MoortgatPick:2005cw}, we adopted $(P_{e^+},P_{e^-})=(+0.6,-0.8)$ and $(+0.3,-0.8)$ in this study. The vertical dashed lines indicate the proposed energies of each lepton collider. In figure \ref{fig3} (left), the distributions $e^+e^-\rightarrow Zh^0\gamma$ are plotted, while on (right) the same distributions for  $e^+e^-\rightarrow ZH^0\gamma$ are shown. Since the $h^0$-state behaves like the SM Higgs boson in the exact alignment limit ($\sba=1$), the distributions in figure \ref{fig3} (left) are in good agreement with the SM tree-level results presented in refs. \cite{Liu:2013cla, Alam:2017hkf}. In this limit, results are similar with the SM results as expected. 
\begin{widetext}
\onecolumngrid
\begin{figure}[htbp]
\centering
\includegraphics[width=0.48\textwidth]{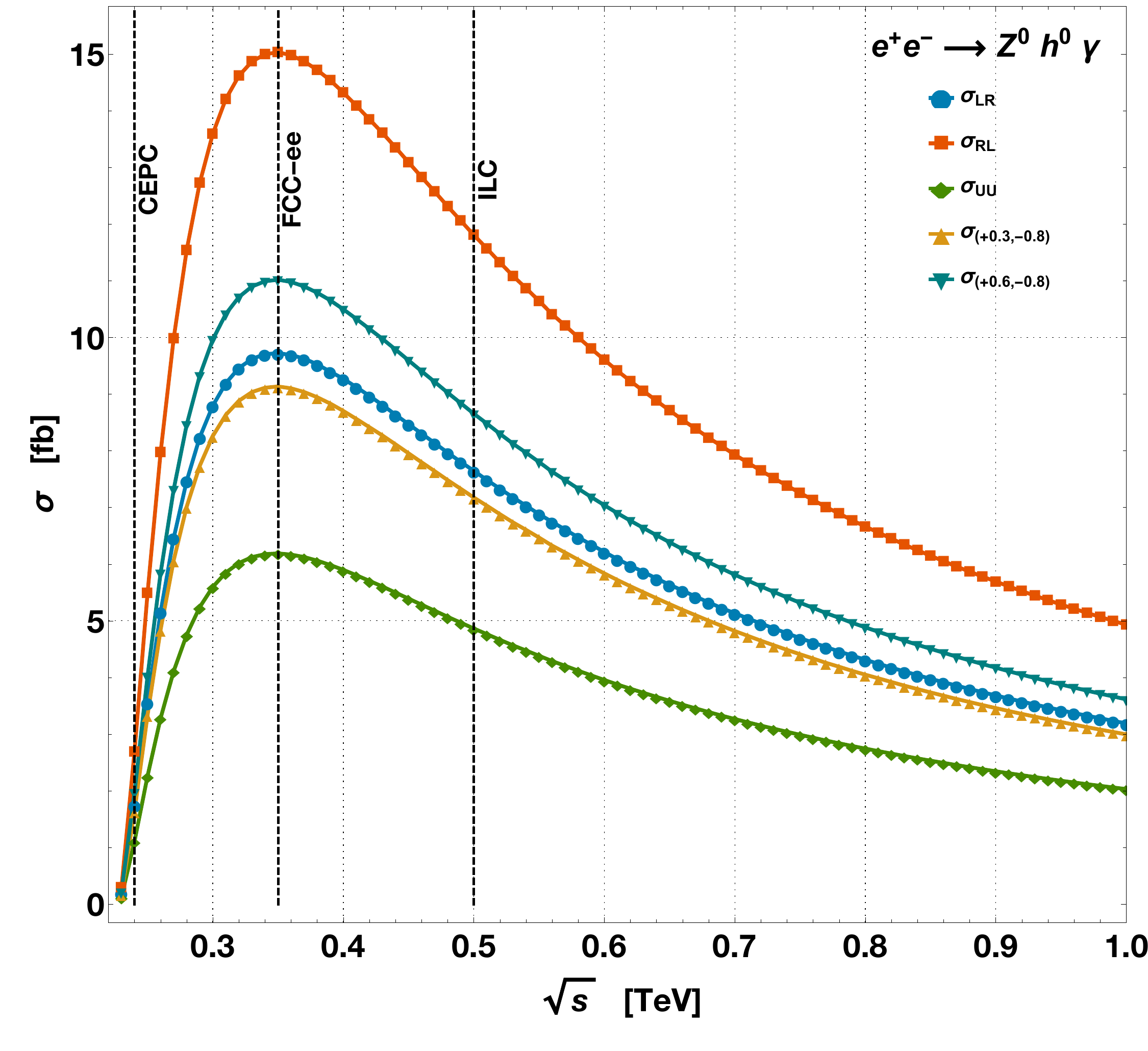}\hfill
\includegraphics[width=0.48\textwidth]{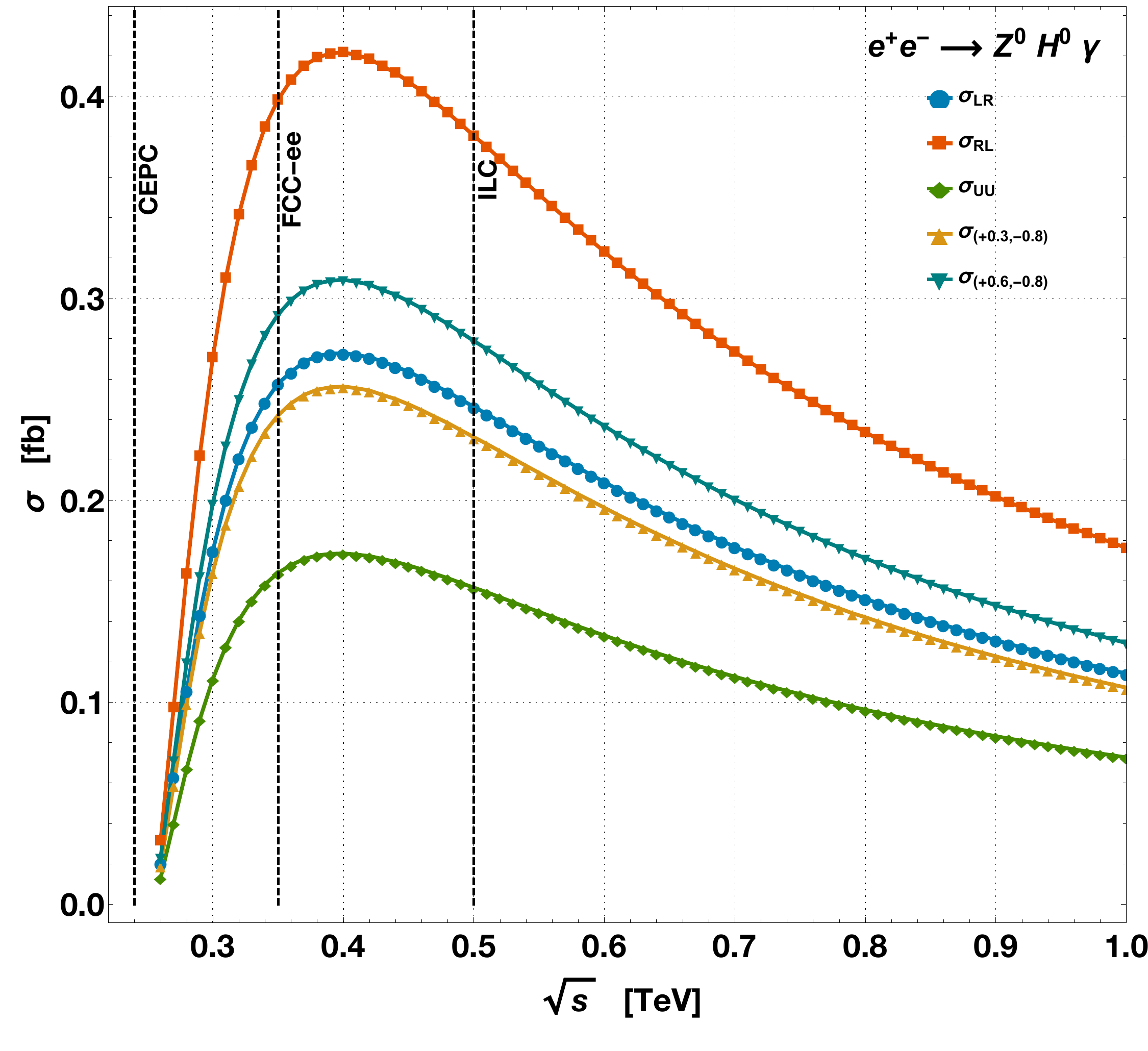}
\caption{
\label{fig3} The distributions of the cross-section are plotted for various polarization configurations of the incoming $e^+e^-$ beams. The calculation is carried for $p_{T,\text{cut}}^\gamma=10\gev$ on both figures. 
        (left): The cross-section distributions are for $\processBir$. The exact alignment limit $\sba=1$ is assumed.
        (right): The cross-section distributions are for $\processIki$. The mixing angle of the neutral Higgs bosons is taken as $\cba=0.2$ ($\sba\approx 0.98$).}
\end{figure}
\end{widetext}
The unpolarized cross-section reaches up to $6.19 \fb$ around the planned collision energy of the FCC-ee collider ($\sqrt{s}=350\gev$), and it slowly declines at higher $\sqrt{s}$ values. 
At the ILC ($\sqrt{s}=500\gev$), the unpolarized cross-section gets $\sigma_{UU}\sim 4.86\fb$. The proposed CEPC running at $\sqrt{s}=240\gev$ could still probe the process, but not as significant as the others, the unpolarized cross-section is around $1.11 \fb$. If the future colliders are capable of polarizing the incoming electron and positron beams, the cross-section will be enhanced dramatically. The polarized cross-sections are also added in both graphics given in figure \ref{fig3}. When the positron beam is completely left-handed and the electron beam is right-handed, $\sigma_{LR}$ is around $9.72 \fb$. In the opposite case where polarization is set to $(+1,-1)$, then $\sigma_{RL}\sim 15.04\fb$. 
Two other polarization configurations are also plotted in figure \ref{fig3} where the peaks are around the FCC-ee operation range,  $\sigma_{(+0.3,-0.8)}\sim 9.14\fb$ and  $\sigma_{(+0.6,-0.8)}\sim 11.02\fb$. We can see that polarizing the beams enhances the cross-section up to a factor of 1.8 in (+0.6, -0.8) configurations.

The mixing angle $\sba$ plays an important role in the calculation, the cross-section of $\processIki$ vanishes in the exact alignment limit ($\sba=1$). Therefore, we let the $\sba$ deviate from unity by only $\sim2\%$, thus the parameter $\cba=0.2$ is assumed in figure \ref{fig3} (right) for the process $\processIki$. The cross-section distributions for possible polarizations of the $e^+e^-$ are presented in figure \ref{fig3} (right). Compared to the previous scattering process, the cross-section of $\processIki$ is smaller. 
The unpolarized cross-section is around $\sigma_{UU}\sim 0.164\fb$ for the planned FCC-ee ($\sqrt{s}=350\gev$) project. 
In addition to this, the cross-section of the completely polarized beams gets $\sigma_{RL}\sim 0.398\fb$ and $\sigma_{LR}\sim 0.258\fb$ at $\sqrt{s}=350\gev$. 
The distributions of the cross-section as a function of energy have the same trend as the previous process. The cross-section peaks around $\sqrt{s}\sim 400\gev$, then it decreases rapidly moving to higher energies due to the suppression of 1/s. 
Due to the kinematical threshold of the process (the total mass of $H^0$-boson, Z-boson, and $\gamma$ with the transverse energy of 10 GeV), the CEPC is only capable of measuring the process $\processIki$ with $m_{h^0}\lesssim m_{H^0}\lesssim 140\gev$ mass constraint. The CEPC has limits measuring this particular parameter space. 

For the ILC ($\sqrt{s}=500\gev$), the cross-section reaches $\sigma_{UU}\sim0.157\fb$. The polarized cross-section values reach up to $\sigma(+0.3,-0.8)=0.231\fb$ and $\sigma(+0.6,-0.8)=0.279\fb$. The polarization of the $e^+e^-$-beams raises the cross-section by a factor of 1.8  compared to the unpolarized values.

\begin{figure}[htbp]
\centering
\includegraphics[width=0.45\textwidth]{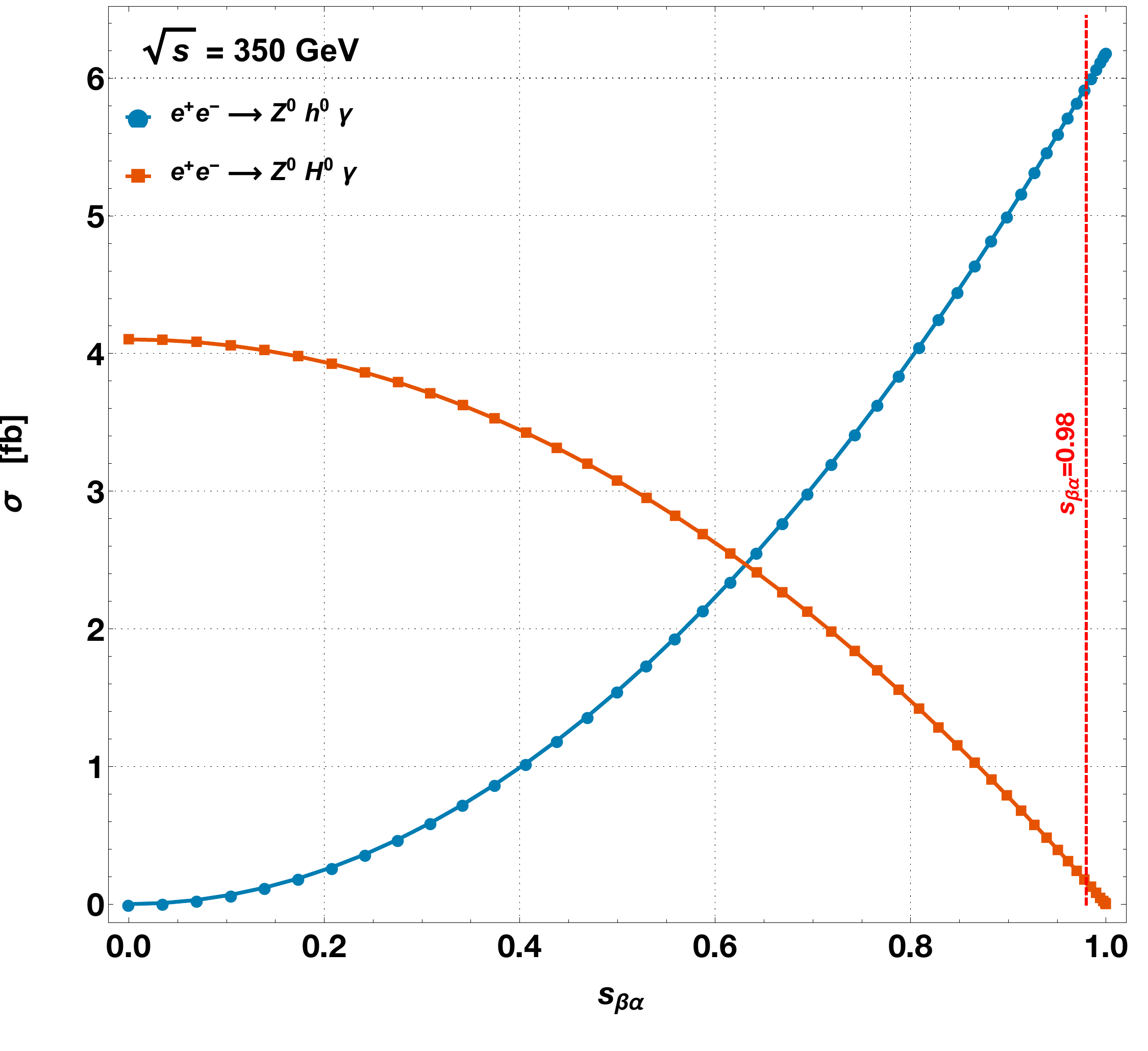}
\caption{
\label{fig4} The cross-section distributions of $\processBir$ and $\processIki$ are plotted as a function of $\sba$ at $\sqrt{s}=350\gev$. The figure shows the cross-sections in departing from the exact alignment limit. The calculation is carried for $p_{T,\text{cut}}^\gamma=10\gev$ and $\tb=10$ on both processes.
        }
\end{figure}

The readers might wonder how the production cross-section of these two processes differ when one departs from the exact alignment limit ($\sba=1$). In figure \ref{fig4}, the distributions of $\processBir$ and $\processIki$ at $\sqrt{s}=350\gev$ are plotted for the unpolarized incoming beams. The vertical dashed line points the $\sba=0.98$ value. The scattering process $\processBir$ gets lower while $\sba\rightarrow0$, and $\sigma(\processBir)$ is $\sim 5.92\fb$ at $\sba=0.98$. It is lowered by $4.36\%$ compared to $\sba=1$. 
Contrary, the process $\processIki$ gets higher for lower $\sba$ values as expected. Deviating from the exact alignment limit by $\sim2\%$ enhances the cross-section; it is $\sigma(\processIki)\sim 0.18\fb$ at $\sba=0.98$.

In table \ref{tab2}, the cross-section values of the $\processBir$ and the $\processIki$ at different CM energies (and colliders) are given for different cuts on the photon's transverse momentum ($p_{T,\text{cut}}^\gamma$). It can be seen that raising the variable $p_{T,\text{cut}}^\gamma$ chops the cross-sections. 
The effect is most dramatic for the CEPC due to the collider's low CM energy. If the photon's transverse momentum is set as low as $p_{T,\text{cut}}^\gamma=5\gev$ the cross-section becomes as high as 3.45 fb for the $\processBir$, unfortunately the $\processIki$ is not available kinematically. Since the $p_{T,\text{cut}}^\gamma$ is a specific parameter of the measuring apparatus, an electromagnetic calorimeter with low energy resolution will increase the acceptance, and this will result in a higher number of events. Considering the other lepton collider proposals, FCC-ee and ILC, the cut on the $p_{T}^\gamma$ is not vital, but raising the cut lowers the cross-section as expected.

\begin{tiny}
\begin{table}[htbp]
\centering 
\caption{\label{tab2} The unpolarized cross-section for $\processBir$ and $\processIki$ processes with varying cuts on photon's transverse energy ($p_{T,\text{cut}}^\gamma$). All energies are given in GeV.}
\footnotesize
\begin{tabular*}{90mm}{c@{\extracolsep{\fill}}| ccc | ccc}
\toprule
Collider    & CEPC    & FCC-ee    & ILC     & CEPC    & FCC-ee    & ILC        \\
$\sqrt{s}$            &  (240)     &  (350)       &  (500)   &  (240)    &  (350)        &  (500)     \\ \hline\hline
$p_{T,\text{cut}}^\gamma$          
                         & \multicolumn{3}{|c|}{$\sigma_{\processBir}$ (fb)}       
                                                & \multicolumn{3}{c}{$\sigma_{\processIki}$ (fb)}    
                                                                            \\ \cline{2-7}
         5                  & 3.450     & 9.189    & 6.536    & -    & 0.266   & 0.223        \\
        10                 & 1.077     & 5.948    & 4.674    & -    & 0.164   & 0.157        \\
        15                 & 0.288     & 4.348    & 3.714    & -    & 0.115   & 0.123        \\
        20                 & 0.027     & 3.345    & 3.091    & -    & 0.085    & 0.101        \\
        25                 & -             & 2.644    & 2.641    & -    & 0.064    & 0.085           \\
\hline
\end{tabular*}
\end{table}
\end{tiny}

The future lepton colliders could be used to study these two scattering processes. If the total luminosity of the future lepton colliders is expected to be at the order of 2000 $\text{fb}^{-1}$ \cite{Fujii:2018mli}, then according to Table \ref{tab2} the total number of events at $\sqrt{s}=350\gev$ (FCC-ee) with $p_{T}^\gamma>10\gev$ are expected to be $\approx 11.9\text{k}$ and  $\approx 330$ for $\processBir$ and $\processIki$ processes, respectively. 
The process $\processBir$ has a remarkable cross-section, but the process $\processIki$ has a considerably smaller one. The branching ratios of the final state particles and the acceptance of the detector are necessary to estimate the precise number of events that can be expected in the detector. Therefore, detector simulations for signal and background channels are required. The process $\processIki$ can be used to study machine learning algorithms due to the expected small number of events.

Lastly, the cross-section of $\processUc$ is at order of $10^{-13}\fb$ due to missing $c_{ZZA^0}$ coupling.  The process cannot be studied in the future lepton colliders. Therefore, the results are not presented. 

\subsection{Polarization of the incoming beams}

\begin{figure}[htbp]
\centering
\includegraphics[width=0.48\textwidth]{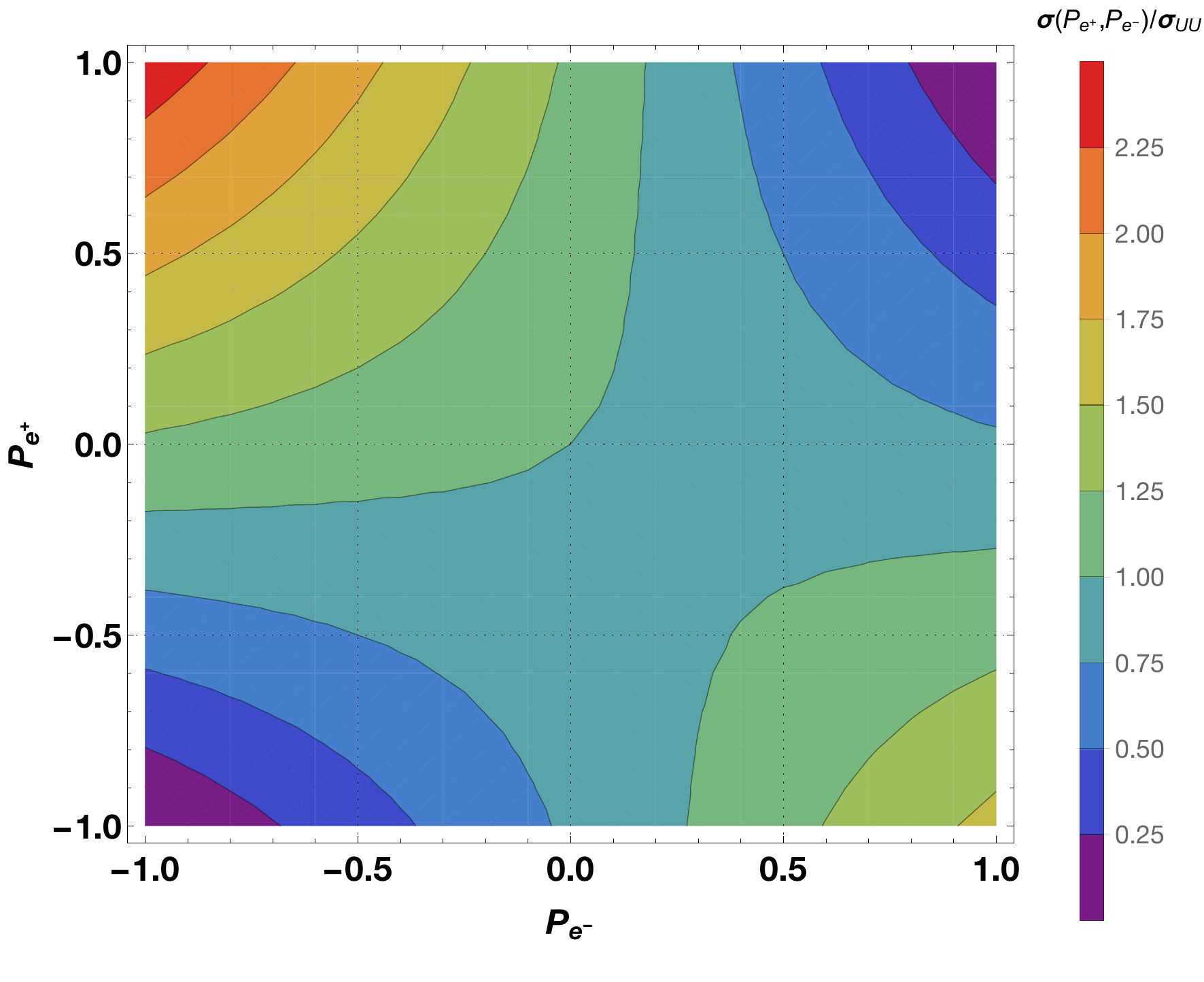}
\caption{\label{fig5} The ratio of the cross-sections ($\sigma(P_{e^+},P_{e^-})/\sigma_{UU}$) as a function of the polarization of $e^+e^-$ beams. The computation assumes $p_{T,\text{cut}}^\gamma=10\gev$, $\tb=10$, $\sba=0.98$ and $\sqrt{s}=500\gev$.
}
\end{figure}

The cross-section contours are given in figure \ref{fig5} for possible polarization configurations of the incoming beams. The ratio of $\sigma(P_{e^+},P_{e^-})/\sigma_{UU}$ is plotted as a function of ($P_{e^+},P_{e^-}$) for the process $\processBir$ at $\sqrt{s}=500\gev$. The cross-section is obtained with $p_{T,\text{cut}}^\gamma=10\gev$, $\tb=10$, $\sba=0.98$ and $m_{H^0}=150\gev$. The same distribution is obtained for the $\processIki$ because the similar Feynman diagrams contribute to the total amplitude.
If the future colliders are able to polarize the positron beam right-handed and electron beam left-handed, then the cross-section values will be higher. It can be seen that in figure \ref{fig5}, the enhancement is raised to 2.25 at the left top corner for both processes. Additionally, the computations showed that the ratio  $\sigma(P_{e^+},P_{e^-})/\sigma_{UU}$ doesn't change with the parameter $\sba$. The contours remain universal because the ratio $\sigma(P_{e^+},P_{e^-})/\sigma_{UU}$ is not a function of energy. Thus, the ratio distribution as a function of the polarization of incoming beams stay unchanged at higher CM energies. 

\subsection{Asymmetry distributions}

We explored the forward-backward asymmetry distribution between the neutral Higgs boson and the photon. 
Since the weak interaction violates parity symmetry, it produces asymmetric contributions to angular observables. Therefore, $A_{\text{FB}}$ could be a good observable for testing type of electroweak model for the Higgs bosons. The asymmetry for each processes is defined as follows:
\begin{equation}
A_{FB}=\frac{\int_0^1\frac{d\sigma}{d\cos\theta_{H_i\gamma}}d\cos\theta_{H_i\gamma}-\int_{-1}^0\frac{d\sigma}{d\cos\theta_{H_i\gamma}}d\cos\theta_{H_i\gamma}}{\int_0^1\frac{d\sigma}{d\cos\theta_{H_i\gamma}}d\cos\theta_{H_i\gamma}+\int_{-1}^0\frac{d\sigma}{d\cos\theta_{H_i\gamma}}d\cos\theta_{H_i\gamma}}.
\end{equation}
As usual the $p_T^\gamma>10\gev$ cut is applied in the calculation. In figure \ref{fig6} (left), the asymmetry is plotted as a function of CM energies for unpolarized incoming beams, free parameters are given in the caption. In the computation, we obtained that in both processes most of the events are accumulated more in the backward direction than the forward direction. As a result, the forward-backward asymmetry distributions become negative for both processes. It gets a value of -0.38 around $\sqrt{s}=240\gev$ (at the CEPC) for the $\processBir$, and at higher CM energies the asymmetry reaches a peak around $\sqrt{s}=270\gev$ then falls.
The processes $\processBir$ and $\processIki$ share the same trend overall, but the asymmetry in $\processIki$ increases slightly in  higher CM energies due to the kinematics of the final state particles. 
The asymmetry at $\sqrt{s}=1\tev$ is obtained to be close to -0.490 and -0.456 for $\processBir$ and $\processIki$, respectively. The asymmetry is calculated for $\sba=0.98$, but it should be noted that the distributions do not change with $\sba$. The dependence on $\sba$ in cross-sections and couplings is dropped in the asymmetry distributions. The asymmetry is dependent on the masses of the Higgses in the final state. In higher $m_{H^0}$ values, asymmetry distribution of $\processIki$ moves to the right with the same trend. 
Various asymmetry distributions for the SM were presented in ref. \cite{Alam:2017hkf}, the same outcome is obtained; the number of events collected in the backward direction is higher than the forward direction. Therefore, the asymmetry has to be negative. However, the numbers are positive in ref. \cite{Alam:2017hkf}, we assume there is a sign error in the calculation. 



The difference between the asymmetry distributions of $\processBir$ and $\processIki$ are calculated due to phenomenological curiosity. The difference in asymmetry between two processes might be used to test whether both processes share similar Feynman diagrams topology. The asymmetry difference as a percentage is defined as follows:
\begin{equation}
\Delta A_{FB}= 100 \times\frac{(A_{\text{FB}}^{\processBir}-A_{\text{FB}}^{\processIki})}{A_{\text{FB}}^{\processBir}}.
 \end{equation}
The percentage of difference in asymmetry between two processes is plotted in figure \ref{fig6} (right) as a function of the CM energy with various  $m_{H^0}$ masses. It can be seen that the difference is steady for $m_{H^0}=125\gev$ in $0.5-1$ TeV range. 
Since the asymmetry for both of the processes has the same tendency as a function of the CM energy, and moves to the right with increasing $m_{H^0}$, the difference rises at high CM energies with increasing $m_{H^0}$ values. It should be noted that the cross-section lowers dramatically with higher $m_{H^0}$. The difference in the asymmetry is around 3.6\% at the FCC-ee, and it quickly reaches to 6\% at the ILC energies for $m_{H^0}=150\gev$. The figure shows that the ILC is more advantageous for measuring the asymmetry difference in various $m_{H^0}$ scenarios.

\begin{widetext}
\onecolumngrid
\begin{figure}[htbp]
\centering
\includegraphics[width=0.48\textwidth]{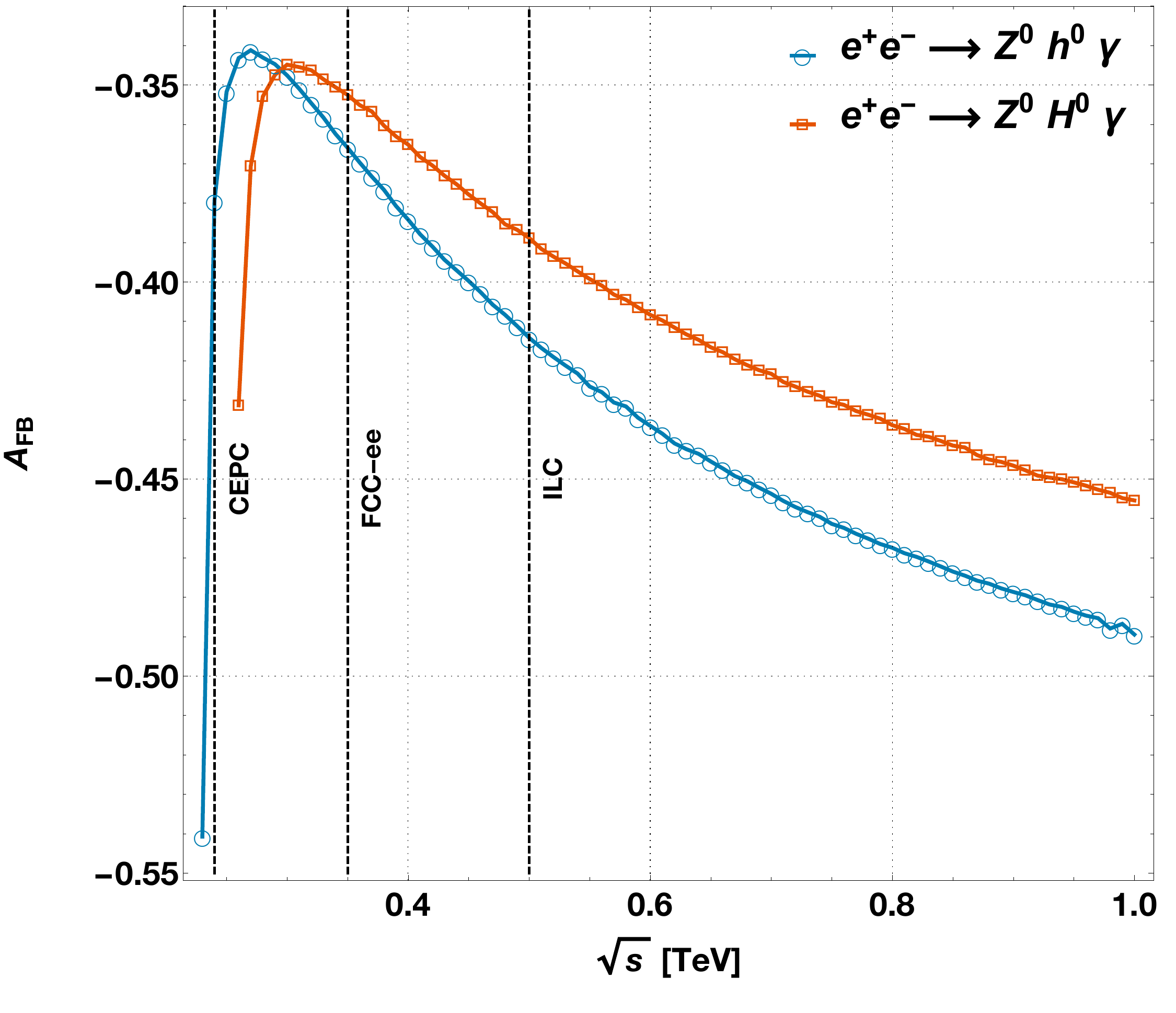}\hfill
\includegraphics[width=0.48\textwidth]{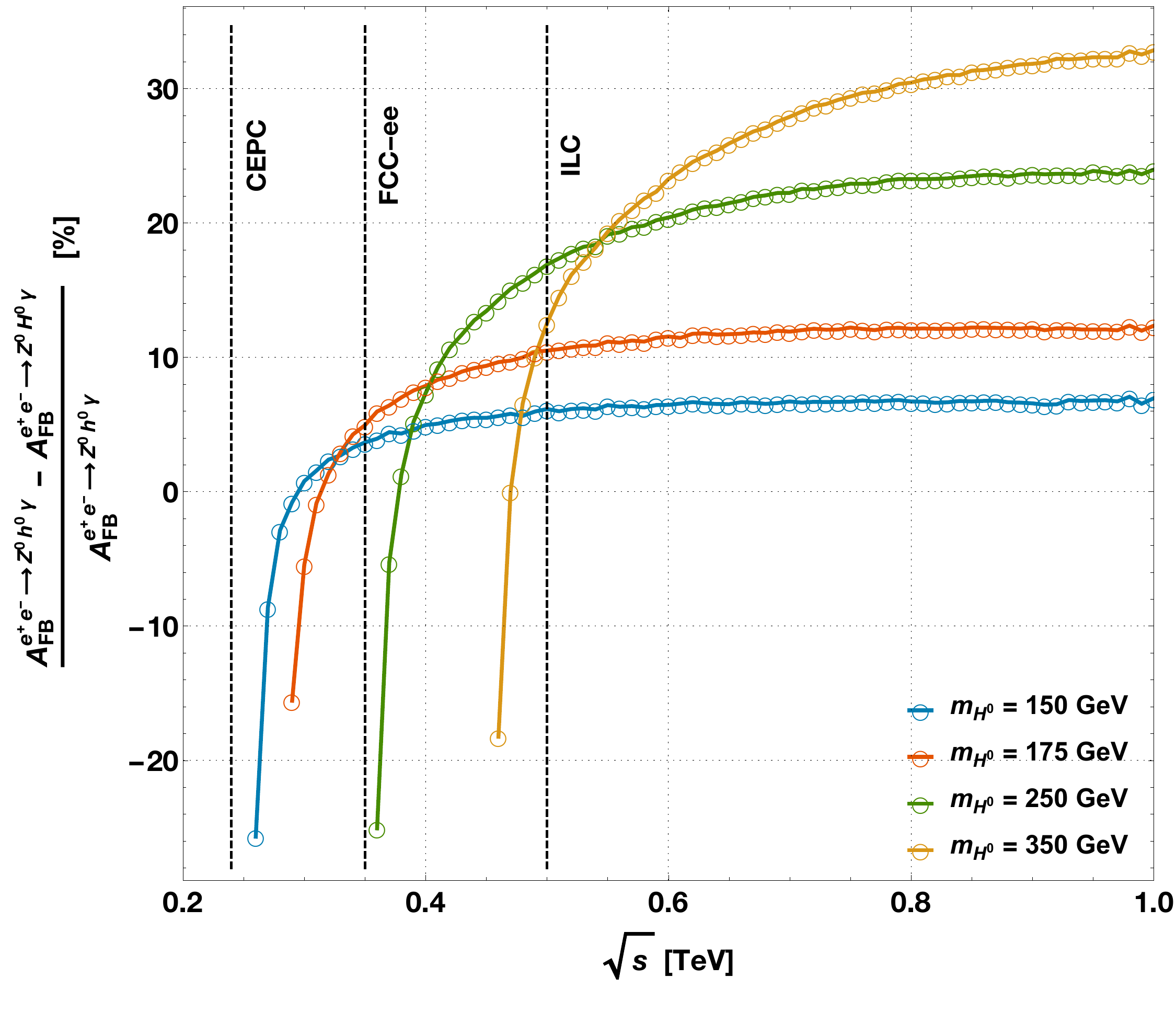}
\caption{\label{fig6} The asymmetry is plotted with $p_{T,\text{cut}}^\gamma=10\gev$, and $\tb=10$ for both processes. $\sba=0.98$ is considered for each processes, but it has no impact on the asymmetry.
(left): The asymmetry distributions with and $m_{H^0}=150\gev$.
(right): The asymmetry difference between two processes as a percentage, ($100\times(A_{\text{FB}}^{\processBir}-A_{\text{FB}}^{\processIki})/A_{\text{FB}}^{\processBir}$) with various $m_{H^0}$ masses.
}
\end{figure}
\end{widetext}


\section{Conclusion}
\label{sec5}

In this study, we focused on computing the scattering processes $e^+e^-\rightarrow Z H_i \gamma$ (i=1,..,3) at the tree level. They were investigated for polarized and unpolarized incoming beams in the context of 2HDM. The results were obtained for a scenario that is favored by recent experimental outcomes and also widely accepted. When $\sba=1$, the couplings of the $h^0$ boson become the same with the SM Higgs boson, therefore one of the Higgs bosons ($h^0$) becomes indistinguishable from the SM Higgs boson (H); this limit is called the exact alignment limit. The cross-section of $\processIki$ diminished completely. No conclusion can be drawn about the 2HDM or the underlying scalar sector from $\processBir$ because it has the same result as the SM.
The production rates of both processes were explored by letting $\sba$ deviated from this limit. Even a small deviation raised the cross-section of process $\processIki$, while the cross-section of $\processBir$ decreased by more than $4\%$.
Since the cross-section distribution of $\processBir$ was a mere function of $\sba$, measuring the cross-section of $\processBir$ is useful to extract the parameter $\sba$. Thus, the couplings $c_{[h^0, H^0]A^0Z}$ and $c_{[h^0, H^0]ZZ}$ can be determined. 
The scattering process $\processIki$ is a function of the same couplings, so its cross-section can be obtained. 
It could be argued that the Yukawa couplings too take part in the calculation, but they are very small to make a noticeable contribution. Therefore, their contribution could be ignored. 

Polarizing the incoming beams changes the contribution of various Feynman diagrams given in figure \ref{fig1}, and the cross-section can get higher values. We compared the polarization configurations, and obtained that the cross-section was enhanced up to a factor of 1.8 with $P_{e^-}=-0.80$ and $P_{e^+}=+0.60$ beams. In conclusion, the polarization of the incoming beams increased the production cross-section and the number of events in the collider. 
The forward-backward asymmetry of the Higgses and the photon was presented for both processes. We obtained that the parameter $\sba$ did not have an impact on the asymmetry distributions. Therefore, it can be deduced that the asymmetry distributions alone can not indicate the underlying model in nature.

Among three proposals, the FCC-ee was the one with a higher cross-section for the $\theprocess$.
The CEPC had low CM energy, but it still had the potential to study the process $\processBir$, but not $\processIki$. The ILC shared the same remarks with the FCC-ee. 
The leading motivations of these future colliders are to measure the properties of the Higgs boson and obtain hints of the extended Higgs sector. This study presented the potential of exploring these processes in future lepton colliders. If these proposed experiments can find deviations from the SM or solid experimental proof of the existence of new physics, a new era will be opened in the history of mankind, and our understanding of nature will change completely.

\acknowledgments

The numerical calculations reported in this paper were partially performed at TUBITAK ULAKBIM, High Performance and Grid Computing Center (\textsc{TRUBA} resources) and the computing resource of \textsc{FENCLUSTER} (Faculty of Science, Ege University). Ege University supports this work, project number 17-FEN-054. 

\bibliography{template-8s_revtex}

\end{document}